\documentclass[aps,footinbib,showpacs,lengthcheck,twocolumn,superscriptaddress]{revtex4-1}
\usepackage{graphicx}
\usepackage{amsmath}
\usepackage{amssymb}
\usepackage{epstopdf}
\usepackage{amstext}
\usepackage{amsthm}
\usepackage{amsfonts}
\usepackage{color}
 \usepackage{braket}

\usepackage[unicode]{hyperref}
\usepackage{microtype}
\hypersetup{
	pdftitle={two rings},
	pdfauthor={},
	pdfsubject={},
	colorlinks=true,
	linkcolor=blue,
	citecolor=blue,
	filecolor=black,
	urlcolor=blue
}

\begin{document}

\title{Hysteresis loops in spinor BECs subject to synthetic gauge fields  }

\author{Shuji Jia}
\affiliation{
	School of Science, Xi'an University of Posts and Telecommunications, Xi'an, China
}	
\author{Jintao Xu}
\affiliation{
	School of Science, Xi'an University of Posts and Telecommunications, Xi'an, China
}	
\author{Qian Jia}
\affiliation{
	School of Science, Xi'an University of Posts and Telecommunications, Xi'an, China
}	
\author{Haibo Qiu}
\email{phyqiu@gmail.com}
\affiliation{
	School of Science, Xi'an University of Posts and Telecommunications, Xi'an, China
}
\author{Antonio Mu\~{n}oz Mateo}
\email{ammateo@ull.edu.es}
\affiliation{Departamento de F\'isica, Universidad de La Laguna, Tenerife 38200, Spain}

\begin{abstract}
	We explore the  hysteretic dynamics of spinor Bose-Einstein condensates of ultracold atoms loaded in static 2D ring geometries and subjected to varying synthetic magnetic fields. Electrically neutral, pseudo-spin-$1/2$ condensates are probed by one-component weak-link potentials that make available paths for the transit of vortices into and out of the ring in both components, and thus can control the number of flux quanta threading the ring hole. For perpendicular fields, domain walls that are dynamically generated in the relative phase of the spin components are shown to play a key role in driving the hysteretic behaviour of the whole system through changes in the net magnetic flux. In the presence of spin-orbit-coupling, hysteresis is exhibited around the phase transitions found by the metastable current states for varying fields.
\end{abstract}

\maketitle

\section[intro]{Introduction}

Artificial gauge fields provide ultracold-gas experiments with a versatile tool to explore the realm of matter-light interactions \cite{Lin2009,Lin2011}. Laser fields can interact with neutral  atoms in similar ways as electromagnetic fields do with electrically charged particles \cite{Galitski2019}.
Below the transition temperature for bosonic atoms to become a Bose-Einstein condensate (BEC), the synthetic electromagnetic interaction allows for the simulation of typical phenomena found in systems characterized by an order parameter, such as superconductors or ferromagnetic materials. Among them, we focus on
hysteresis (magnetic hysteresis for the sake of comparison), that is, the phenomenon of a path dependent, nonlinear magnetization, as the order parameter determined by spin alignment, in response to an external magnetic field \cite{Kittel}.

Hysteresis loops result from magnetization paths that are doubly connected, where each path is associated with a corresponding metastable state of the system. This phenomenon finds a close analog in the response of a BEC wave function (as the order parameter) to a synthetic magnetic field \cite{Garcia2001,Mueller2002,Baharian2013}, where the condensate angular momentum, in a configuration threaded by the field, becomes the counterpart of the spin (plus orbital) angular momentum of the ferromagnetic material.
 The first realizations of this type made use of BECs loaded into rotating ring traps \cite{Wright2013,Eckel2014}, whose order parameter transits across different persistent-current states by means of a weak link potential that breaks the rotational symmetry. These systems can be used as basic units of potential atomtronic circuits, and are featured by their particle current-phase relationship in analogy with the usual electrical current-voltage response \cite{Eckel2014b}.  
 
 In this work, we study the  hysteretic dynamics of pseudo-spin-$1/2$ Bose-Einstein condensates loaded in static 2D ring geometries in the presence of synthetic magnetic fields. Our analysis  begins with revisiting scalar condensates subjected to varying weak-link potentials (the static equivalent of the rotating condensates \cite{Wright2013}), and proceeds with
 spinor systems in the presence of either magnetic fields that are perpendicular to the ring plane \cite{Lin2009}, or in-plane magnetic fields akin to the usual spin-orbit coupling with equal contributions of Rashba and Dresselhaus terms \cite{Lin2011}. 
  For both types of fields, our case study focuses on weak links that are restricted to operate in just one spin component, which introduces a distinct feature with respect to scalar systems. From the latter systems (see e.g. Refs. \cite{Kanamoto2008,Mateo2015,Zhang2016}), it is known that solitonic states that include dark solitons or solitonic vortices provide an energy barrier that preserve the metastable states. For spinor condensates we demonstrate the comparable role played by domain walls of the relative phase \cite{Son2002}, which, besides, are key agents in the generation of hysteresis loops by the whole system. These domain walls link vortices in different spin components, and drag them into or out of the rings; in this way, for varying magnetic fields, they switch on and off the particle currents in the ring bulk, and increase or decrease by one unit the number of flux quanta threading the ring hole. In the presence of in-plane fields, which produce vanishing flux, hysteresis is exhibited around phase transitions, and it is shown that can be supported by half-vortex states.  Our results are based on experimentally feasible systems, and obtained from numerical simulations of the 2D Gross-Pitaevskii equation.

This paper is structured as follows: Section \ref{sec:model} introduces the mean-field model used to study the effects of synthetic magnetic fields in ring-shaped condensates, along with the relation between vortex currents and magnetic flux. Section \ref{sec:hysteresis} discusses the formation of hysteresis loops for varying magnetic fields, first in scalar condensates, and later in spin-1/2 condensates; in the latter case, both regular and spin-orbit-coupling gauge fields are considered. Section \ref{sec:conclusions} summarizes our results and presents prospects of related work.

\section[model]{Model}
\label{sec:model}

We consider a 2D  Bose-condensed system of electrically-neutral particles, moving in the $x-y$ plane, in the presence of an artificial gauge potential ${\bf A}\equiv(A_x,\,A_y)$ chosen to be either ${\bf A}=(-B_z\,y,\,0 )\,I_2$,  or ${\bf A}=(0,\,-B_x\,\ell_\nu )\,\sigma_z$, as realized in Refs. \cite{Lin2009,Lin2011}, where $B_z$ and $B_x$ are constant field strengths, $I_2$ and $\sigma_z$ are the $2\times 2$ identity matrix and the $z$-Pauli matrix, respectively, and $\ell_\nu=\sqrt{\hbar^2/(4M\nu)}$ is the characteristic length of the linear coupling $\nu$ in the spinor system (see below). 
 As it is well known (see e.g. \cite{Jackson1999}), the gauge field enters the particle dynamics, in a minimal coupling configuration, through the two-dimensional mechanical momentum of the particles  $\hat{\bf \Pi}=\hat {\bf p}-q{\bf A}$, where $\hat {\bf p}=-i\hbar\nabla$ is the canonical momentum operator, and $q$ is (not the electric charge here but) the strength of the coupling between atoms and gauge field.  The former potential $A_x=-B_z\,y$ is the analogue of the usual Landau gauge in electromagnetism, which gives rise to the cyclotron orbits of  frequency $\omega_c=q|B_z|/M$ in the presence of a perpendicular, constant magnetic field  $B_z$.
 The latter potential $A_y=-\sigma_z\,B_x\,\ell_\nu$ provides a spin-orbit interaction term $(q B_x\,\ell_\nu/M)\, \sigma_z\,\hat p_y$ with equal
 contributions of Rashba and Dresselhaus couplings \cite{Lin2011}; this potential can also be understood as generated in the discrete direction $z$ (spanned by the linear coupling) by an in-plane, constant magnetic field $B_x$, and so the cyclotron frequency is $\omega_c=q|B_x|/M$.

 Pseudo-spin-$1/2$ condensates made of two coupled components, with order parameter,  $\psi=[\psi_\uparrow\,\psi_\downarrow]^T$ have their dynamics ruled by the Hamiltonian [written in the basis of Pauli matrices $(I_2,{\boldsymbol\sigma})$]
 \begin{align}
 \hat H=&\left[\frac{(\hat p_x-qA_x)^2+\hat p_y^2+(qA_y)^2}{2M}+V_R+\frac{gn+\,W}{2}\right]\,I_2\nonumber\\ &\qquad+\left(\frac{q B_x \ell_\nu}{M}\hat p_y+\frac{gn_s+\,W}{2}\right) \,\sigma_z-\nu\sigma_x.
 \label{eq:Hspinor}
 \end{align}
  The interparticle interaction, of strength $g$, splits into a term proportional to the total density $n=|\psi|^2=|\psi_\uparrow|^2+|\psi_\downarrow|^2$, and a term proportional to  the local population imbalance or spin density $n_s=|\psi_\uparrow|^2-|\psi_\downarrow|^2$. 
  $\nu$ is the energy of the linear coupling that allows for the spin flips between the condensate components. 
 The external potential $V({\bf r})$, confining the atomic cloud, consists of two terms $V({\bf r})=V_R({\bf r})+W({\bf r})$: a harmonic potential that imposes a ring geometry of mean radius $R$
\begin{align}
V_R({\bf r})=\frac{M\omega^2}{2} (r-R)^2,
\label{eq:Vring}
\end{align} 
where $r^2=x^2+y^2$, and a Gaussian weak-link potential
\begin{align}
W({\bf r})=W_0\exp\left[-2\frac{ (x-x_0)^2}{s_x^2}-2\frac{ y^2}{s_y^2}\right],
\label{eq:W}
\end{align} 
of height $W_0$, widths $\{s_x,\,s_y\}$, and centered at $(x_0,\,y_0)=(R,0)$ (unless otherwise stated), which acts as a barrier for the superfluid flow only in the spin component $\psi_\uparrow$.
 
The order parameter follows the Gross-Pitaevskii equation
\begin{align}
i\hbar \partial_t \psi =\hat H\psi,
\label{eq:GP}
\end{align}
and fulfills the continuity equation
\begin{align}
 \partial_t |\psi|^2+\partial_x J_x+\partial_y J_y=0,
 \label{eq:continuity}
\end{align}
that leads to the particle-number conservation $\int dx\,dy\,|\psi|^2=N$. Equation (\ref{eq:continuity})
involves the 2D current density ${\bf J}=(J_x,\,J_y)=\Re(\psi^*\hat{\bf \Pi}\psi)/M$, by means of which the superfluid velocity is defined as ${\bf v}={\bf J}/{|\psi|^2}$. 
In addition, the spin density satisfies
\begin{align}
\partial_t n_s+\nabla {\bf J}_s=-\frac{2\nu}{\hbar}\sqrt{n^2-n_s^2}\,\sin\varphi, 
\label{eq:Scontinuity}
\end{align}
where ${\bf J}_s={\bf J}_\uparrow-{\bf J}_\downarrow$ is the spin current density, and $\varphi=\arg\psi_\uparrow-\arg\psi_\downarrow$ is the relative phase. States with in-phase, $\varphi=0$, or out-of-phase, $\varphi=\pi$, spin components suppress spin flips and conserve the total population imbalance $N_s=\int d{\bf r}\,n_s$.

Notice that the synthetic gauge potential, entering $\hat \Pi$ in the equation of motion (\ref{eq:GP}), is not dynamical, i.e., there is no alteration of the underlying field strength due to the wave function dynamics; in contrast to the electromagnetic gauge, this synthetic gauge is fixed in the experiment.

Stationary states are found as space-time coordinate separable wave functions $\psi=\psi(x,y)\exp(-i\mu\, t/\hbar)$, where $\mu$ is the chemical potential.
For low-enough gauge field values, $\omega_c\ll \mu/\hbar$, the ground state is expected to follow a Thomas-Fermi density profile determined by the external potential. For increasing values of the gauge field, beyond a field threshold, approximately $B_{th}\propto \mu^{-1}$ \cite{Lundh1997}, quantum vortices enter the atomic cloud and form part of the ground state \cite{Lin2009}; the resulting vortex lattices for scalar BECs have been reported in the literature, see for instance the numerical simulations of Ref. \cite{Zhao2015}, where the equivalent symmetric gauge potential ${\bf A}'=(-y,\,x)\times B_z/2$ is used along with isotropic harmonic trapping.
 
In what follows, we will make use of non-dimensional quantities, denoted by tildes, that are measured in a system of units determined by the fundamental set $\{\ell,\,M\ell^2/\hbar,\,\hbar^2/(M\ell^2)\}$, made of units of length, time, and energy, respectively, where $\ell/a_{ho}=10^{1/4}$ and $a_{ho}=\sqrt{\hbar/(M\omega)}$.

\subsection{Scalar BECs}

Single-component systems will be described by the scalar BEC wave function $\psi$, and driven by the scalar Hamiltonian $\hat H={\hat{\bf \Pi}^2}/{2M}+V_R+W+g|\psi|^2$.
It is insightful to explicitly recall the stationary states of this Hamiltonian in the absence of weak link. To this end, we use the Thomas-Fermi ansatz, in polar coordinates $(r,\phi)$,
\begin{align}
\psi_{TF}(r,\phi)=\sqrt{\frac{\mu-V_R(r)-\hbar^2m^2/(Mr^2)}{g}}\, e^{i(m\phi-\mu t/\hbar)},
\label{eq:ansatz}
\end{align}
such that $\psi_{TF}=0$ when $\mu<V_R(r)+\hbar^2m^2/(Mr^2)$, where $m=0,\,\pm1,...$ is the phase winding number. This allows us to estimate the typical radial width of the ring  $\Delta R\approx 2\sqrt{2\mu/(M\omega^2)}$,  maximum density $n_{\max}\approx \mu/g$, and  healing length $\xi=\sqrt{\hbar^2/(M\,g \,n_{\max})}$. Notice that current-carrying states, having $m\neq 0$, present depleted density profiles $n_{\max}\approx [\mu-\hbar^2m^2/(MR^2)]/g$.

\begin{figure}[t]
	\centering
	\includegraphics[width=\linewidth]{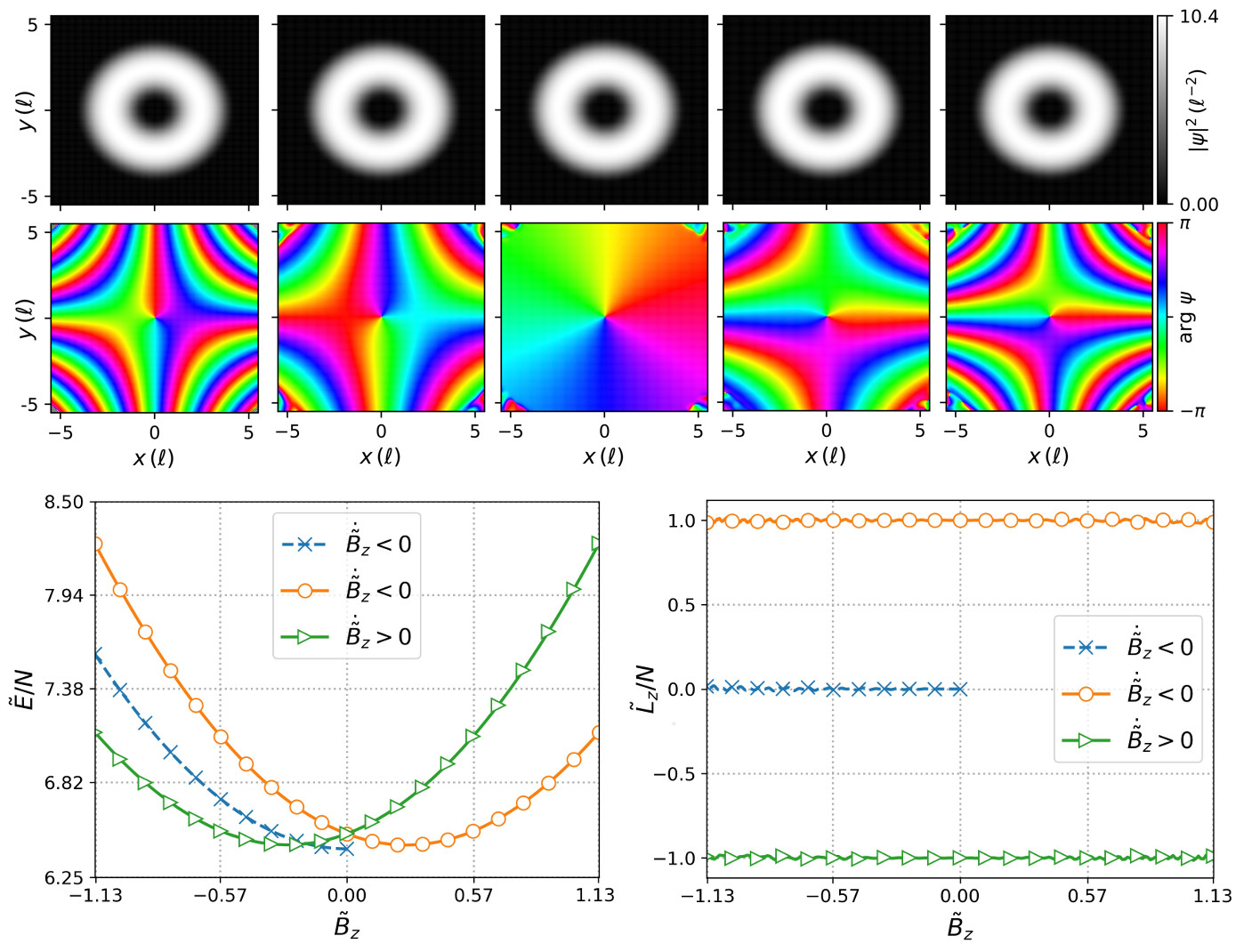}\
	\caption{ Scalar BEC subjected to the adiabatic variation of the synthetic,  perpendicular magnetic field $\tilde B_z=\pm\tilde B_z^0(1-2\, t/T)$ in the absence of weak-link potential. The bottom panels show the energy per particle (left) and canonical angular momentum per particle (right) versus magnetic field. The top panels depict the density and phase profiles at times $t/T=0,\,0.25,\,0.5,\,0.75,\,1$, along the adiabatic path that starts from the ground state at $\tilde B^0_z= 1.1$. } 
	\label{fig:noWL}
\end{figure}
\begin{figure}[t]
	\centering
	\includegraphics[width=\linewidth]{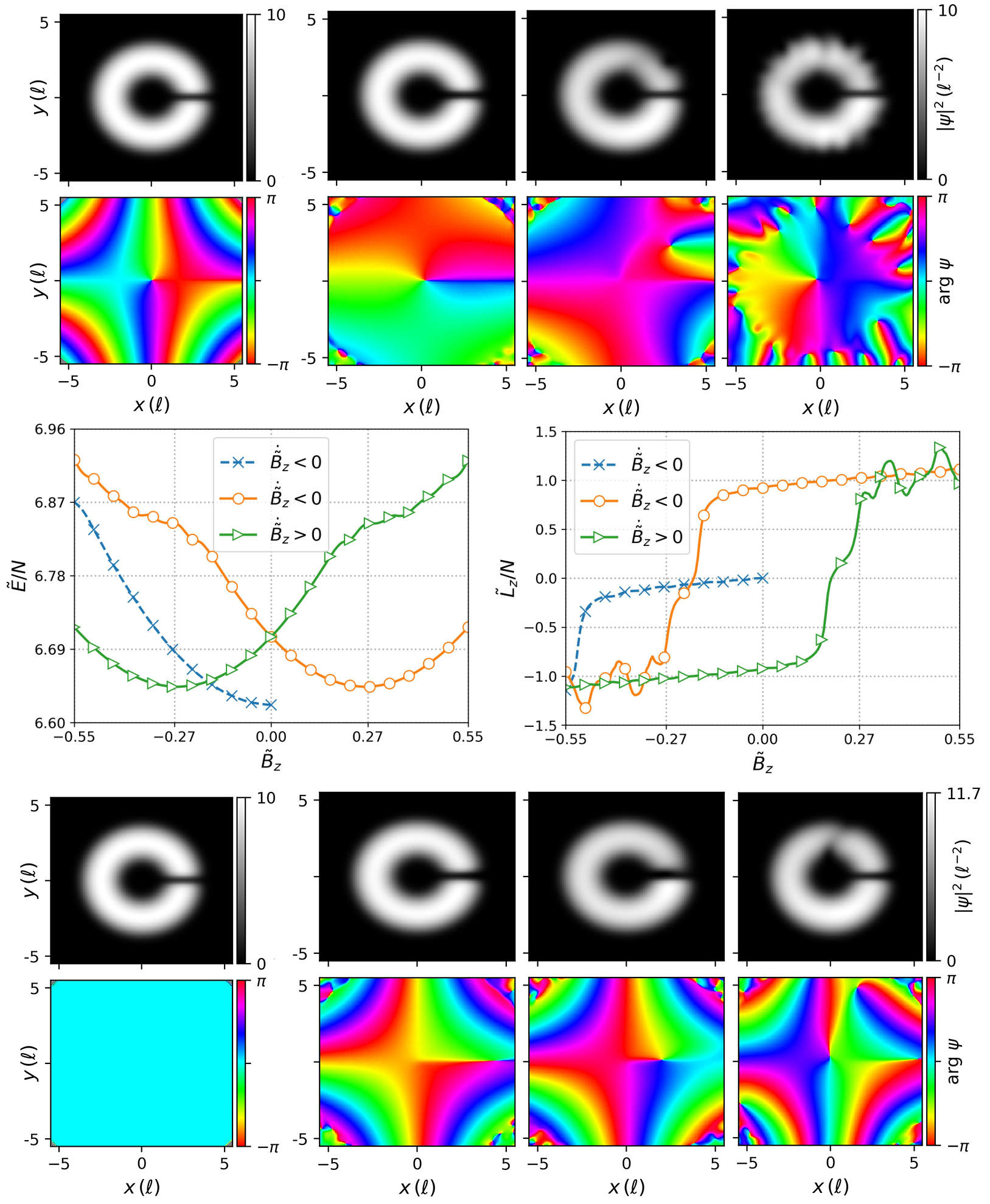}\
	\caption{Same processes as in Fig.~\ref{fig:noWL} but with a weak-link potential. The top panels correspond to times $\tilde B_z=0.55,\,  -0.14,\,  -0.2,\,  -0.33$,  in the path beginning with the ground state at $\tilde B_z^0=0.55$, whereas the bottom snapshots are taken at times $\tilde B_z=0.0 ,\, -0.49,\,  -0.52,\,  -0.55$, along the path that begins  with $\tilde B_z^0=0$.} 
	\label{fig:sWL}
\end{figure}
\subsection{Flux quantization}

In the presence of gauge fields, the circulation of the velocity around a loop, $\Gamma=\oint {\bf v\cdot dl}$, is not quantized. For a scalar condensate
\begin{align}
\Gamma=\frac{1}{M}\oint (\hbar{\bf \nabla \arg \psi}-q{\bf\,A})\cdot {\bf dl}=\frac{2\pi\hbar}{M}\left( \,m-\frac{\Phi}{\Phi_0}\right),
\label{eq:flux}
\end{align}
where $\Phi=\oint{\bf\,A}\cdot {\bf dl}$ is the flux threading the loop, $\Phi_0=2\pi\hbar/q$ is the flux quantum, and $m=0,\pm1,\pm2,...$ is an integer. For loops encircling the origin, $m$ is the winding number in Eq. (\ref{eq:ansatz}), and it is therefore associated with the presence of vortices at the ring center.
In this way, for generic regions encircled by vanishing velocity circulation, flux quantization and vortices are related by Eq. (\ref{eq:flux}), to give  $\Phi=m\, \Phi_0$.
Figures \ref{fig:noWL} and \ref{fig:sWL} show scalar condensates having vortices within the ring hole, thus being threaded by an integer number of magnetic-flux quanta. On the other hand, for loops with  non-zero velocity circulation the fluxoid $\Phi'=\Phi+(M/q)\oint {\bf v\cdot dl}=m\, \Phi_0$ is always quantized \cite{Barone}.

\begin{figure*}[t]
	\centering
	\includegraphics[width=0.8\linewidth]{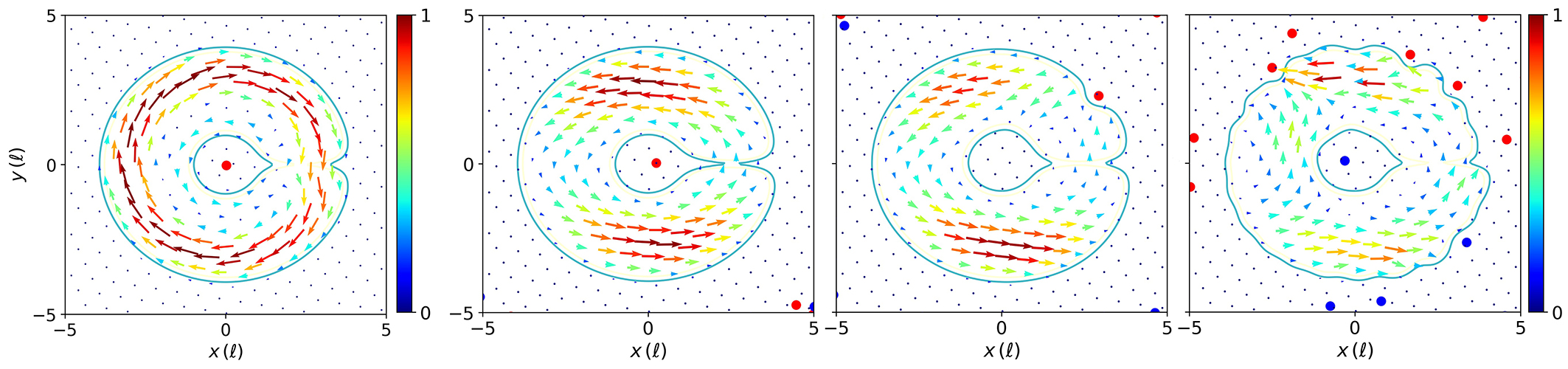}
	\includegraphics[width=0.8\linewidth]{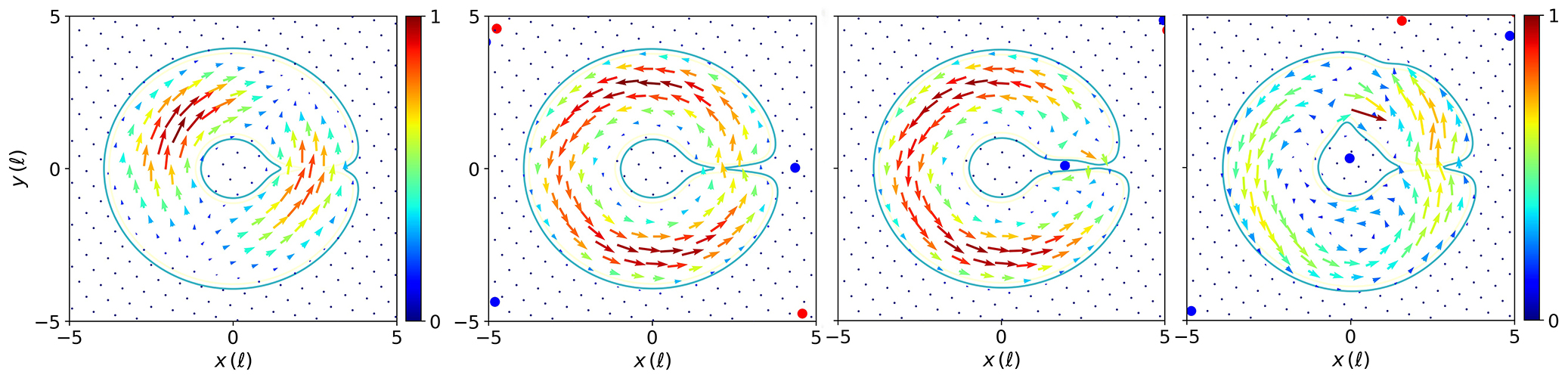}
	\caption{ Same as Fig. \ref{fig:sWL} but showing the current density for paths starting with a vortex  (top row) and without a vortex (bottom row). Vortices and antivortices are indicated by blue and red solid circles, respectively.
	} 
	\label{fig:sWLj}
\end{figure*}

For spin-1/2 condensates, the circulation of the total velocity ${\bf v}$ in the $x-y$ plane gives
\begin{align}
\Gamma=\frac{1}{2M}\oint \left(\hbar{\bf \nabla}\theta-2q{\bf\,A}+\frac{n_s}{n}\hbar\nabla\varphi\right)\cdot {\bf dl}\nonumber \\=
\frac{\pi\hbar}{M}\left(m-2\frac{\Phi}{\Phi_0}+ \frac{1}{2\pi}\oint \frac{n_s}{n}\nabla\varphi\cdot {\bf dl}\right),
\label{eq:Sflux}
\end{align}
where $\theta=\arg\psi_\uparrow+\arg\psi_\downarrow$ is the total phase. Therefore, in general, the extra term in the gradient of the relative phase in Eq. (\ref{eq:Sflux}) precludes  flux quantization. For perpendicular magnetic fields $B_z$, if the condensate components are phase locked, so $\varphi=0$, the same conditions for flux quantization of scalar condensates are recovered, but in units of half the flux quantum $\Phi_0/2$. In this regard, for a loop around the origin, notice that $m=\pm 1$ in Eq. (\ref{eq:Sflux})  is produced by a half vortex state, that is a state having a vortex of charge $m=\pm 1$ in one of the spin components, and no vortex, so $m=0$, in the other component. 

For in-plane magnetic fields, $B_x$, the flux is zero, and Eq. (\ref{eq:Sflux}) predicts the quantization of
the integral on the relative-phase term for loops around which the velocity circulation vanishes; if in addition $n_s/n$ is constant around the loop, it has to match a rational number $n_s/n=m/j$, with $j$ integer.

\section{Hysteresis loops}
\label{sec:hysteresis}

The metastability of the superfluid flow is probed by adiabatically changing the gauge field or, equivalently, the synthetic magnetic field. First, an initial field value is selected along a given spatial direction, say $\pm B_z^0$, and a corresponding stationary state is numerically found, $\psi_0(x,t)=\psi_0(x)\exp(-i\mu t/\hbar)$. Next, this state is subjected to a slow field variation $B_z(t)=B_z^0(1\mp 2\, t/T)$, during a time interval $T\gg \hbar/\mu$, until the initial field value is reversed $B_z(T)=\mp B_z^0$. Different signs of the initial field give rise to distinct paths associated with different metastable states, which, even despite having equal mean energies,  can be distinguished in a chart of canonical angular momentum versus field strength (see the bottom panels of Fig. \ref{fig:noWL}).
Transitions between these paths can be caused by the presence of weak links \cite{Mueller2002,Baharian2013,Mateo2015}, as a result of which the two paths can be connected and the hysteresis loops closed (see the central panels of Fig. \ref{fig:sWL}). Alternative paths starting at zero magnetic field, $B_z(t)=\mp B_z^0\, t/T$, are also followed until $B_z(T)=\mp B_z^0$ (see the blue lines with cross symbols in Figs. \ref{fig:noWL} and \ref{fig:sWL}). As can be seen in Fig. \ref{fig:sWLj}, hysteresis loops are driven by vortices and antivortices (indicated by blue and red circles, respectively). They give rise to complex current patterns in the ring bulk that vanish at the outer and inner borders of the rings; this fact translates into a vanishing velocity circulation along the inner border, which, according to Eq. (\ref{eq:flux}), allows for the calculation of flux quanta by counting vortices within the ring hole.

To analyze these phenomena, we choose systems  
with typical chemical potential $\mu=\sqrt{10} \,\hbar\omega$ confined in a ring of radius $R=25\,\xi$, where a weak link is acting within the Josephson regime (see below) with parameters $W_0=2.5\,\mu$,  $s_y=2/\sqrt{10}\,\xi$, and $s_x= R$, such that a long transverse junction is produced; the typical radial width of the atomic cloud is $\Delta R\approx 3\sqrt{10} \,\xi$. For a feasible experimental realization with  $^{23}$Na atoms, these values would correspond to a BEC of $\sim 2\times 10^4$ particles trapped by a harmonic potential of radial angular frequency $\omega/(2\pi)= 60$ Hz, around an average radius of $R=12 \,\mu$m, and perpendicular angular frequency $\omega_z/(2\pi)= 600$ Hz. Similar spinor rings have been realized in ultracold gas experiments \cite{Beattie2013}.
It is insightful to begin the analysis through the cases of scalar condensates presented before.

\subsection{Hysteresis in scalar condensates}

Phase-slip events allow the system to change the phase winding around the ring by an integer number of $2\pi$ \cite{Anderson1966}, thus to make a transition into a different metastable state.
Figure \ref{fig:noWL} shows the adiabatic paths followed by the ground states obtained at $B_z^0=\pm 1\,\hbar/(\ell^2\,q)$ (i.e. $\tilde B_z^0=\pm 1$) in the absence of weak link. Despite the energy crossing at $B_z=0$, the adiabatic variation preserves the metastable states and Landau-Zener tunneling is prevented by the topological protection of phase winding numbers. Therefore no phase-slips are induced, thus no flux variation is produced; notice that the phase  winding $m$ of the states [$m=+1$ (resp. $-1$) in the orange with circles (green with triangles) path starting at $\tilde B_z^0=1.1$ (resp. $\tilde B_z^0=-1.1$)] is kept all along the paths.

Weak-link potentials make phase slips,  and with them flux variation, possible, which along with the metastability of the superfluid flow can give rise to hysteresis loops. These potentials break the rotational symmetry of the system and obstruct the azimuthal current flow of the particles \cite{Ramanathan2011,Wright2013}, which results in a system with weaker superfluidity \cite{Chauveau2023}. In a one dimensional ring, the weak link is characterized by the potential height $W_0$, and the azimuthal length, $s_y$ in Eq.~(\ref{eq:W}), to be compared with bulk properties, the chemical potential $\mu$ and the healing length $\xi$, respectively. As in superconductors, the Josephson effect (that is, the particle currents  across the weak link that are proportional to the sine of the relative phase) is expected to take place for short junctions, $s_y<\xi$, that are determined by high barriers $W_0\ge \mu$ \cite{Likharev1979,Watanabe2009,Piazza2010}. In the 2D system, the radial width of the condensate $\Delta R$ (hence the radial width of the junction) introduces a second length scale that determines, according to the ratio $\Delta R/\xi$,  the long or short transverse character of the junction \cite{Barone}. Topological defects such as dark solitons \cite{Kanamoto2008,Zhang2016} and vortices \cite{Piazza2013,Mateo2015} are pinned by the weak link and appear in intermediate states during phase slips. 
\begin{figure}
	\centering
	\includegraphics[width=\linewidth]{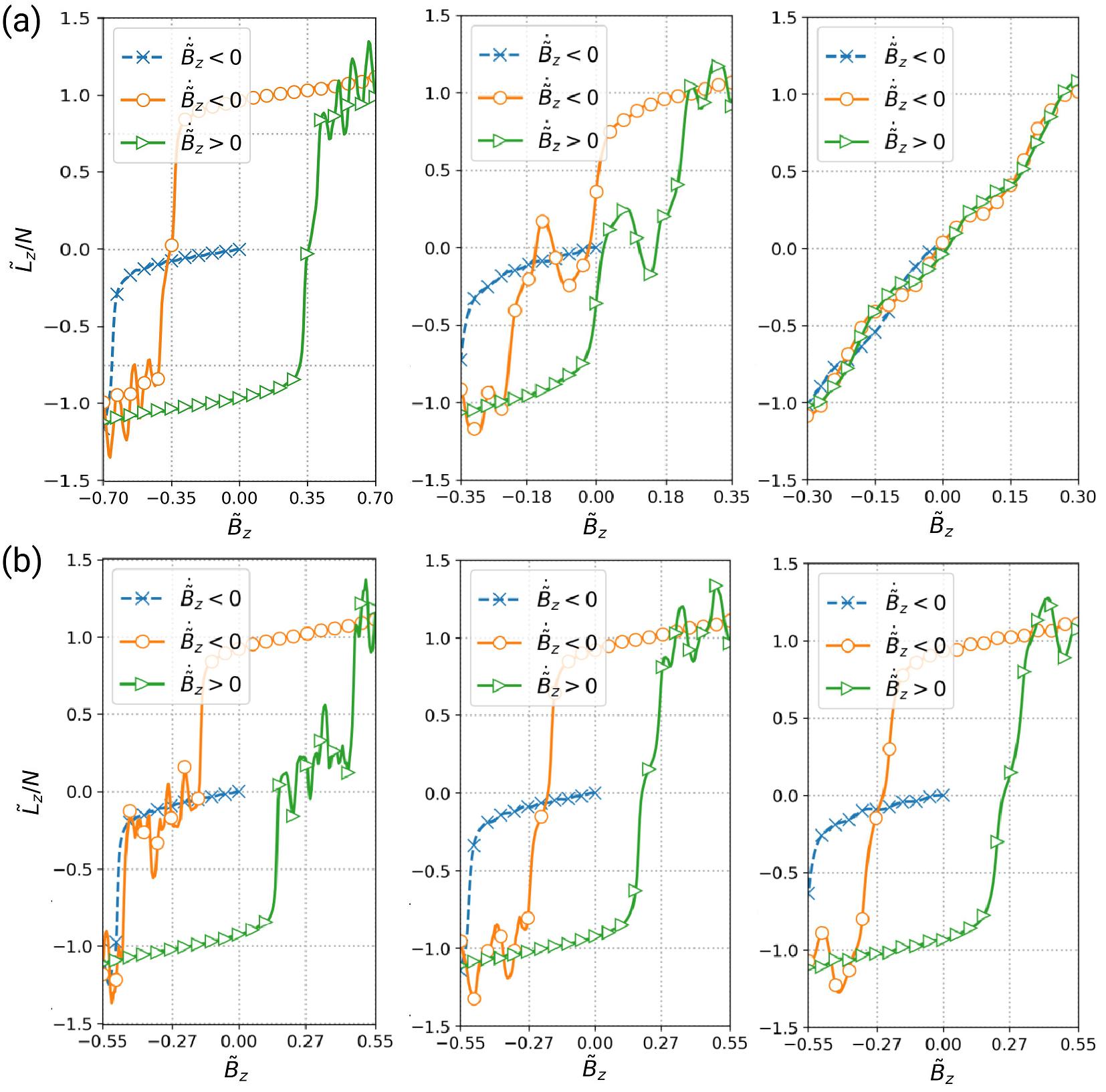}
	\caption{ (a) Hysteresis loops for varying weak link azimuthal  length $s_y$. Left to right: $\tilde s_y= 0.15, 0.3, 0.8 $.
		(b)  Hysteresis loops for different magnetic field ramps. Left to right: $\dot B_z/B_z^0 = 0.018,\, 0.036,\, 0.073$. } 
	\label{fig:WLsy}
\end{figure}

 Figure \ref{fig:sWL} depicts the characteristic hysteretic loops (center-right panel) created in the presence of weak links. In contrast with the paths shown in Fig. \ref{fig:noWL}, the central vortex of the initial state at $\tilde B_z^0=0.55$ (top panels) finds an adiabatic path through the weak link to exit the system, which produces the reduction by one unit of the winding number $m\to m-1$; due to the metastability of the system, this transition is made at values of the magnetic field where the lowest energy state has windineng number $m-2$, and one can see an almost immediate second phase slip $m-1\to m-2$ due to the entry of an antivortex in the ring [see the density and phase snapshots of Fig. \ref{fig:sWL}].
 For increasing gauge field, the opposite way is followed  by the initial antivortex that eventually leaves the ring, producing $m\to m+1$, and the subsequent phase slip $m+1\to m+2$ produced by the entry of a vortex. Similarly, the initial state without vortex at $\tilde B_z^0=0$ (bottom panels)  ends up hosting an antivortex that enters the ring via the weak link. Since there is no dissipation, one cannot see the system settling in the ground state with the corresponding winding number after a phase slip event; the adiabatic processes do not prevent additional vortices in the outer part of the system to modulate the density profile along the ring, storing the extra energy. As a result, oscillations of the angular momentum take place around the smooth path followed by the opposite initial stationary state.

 \begin{figure*}[ht]
 	\centering
 	\includegraphics[width=1.0\linewidth]{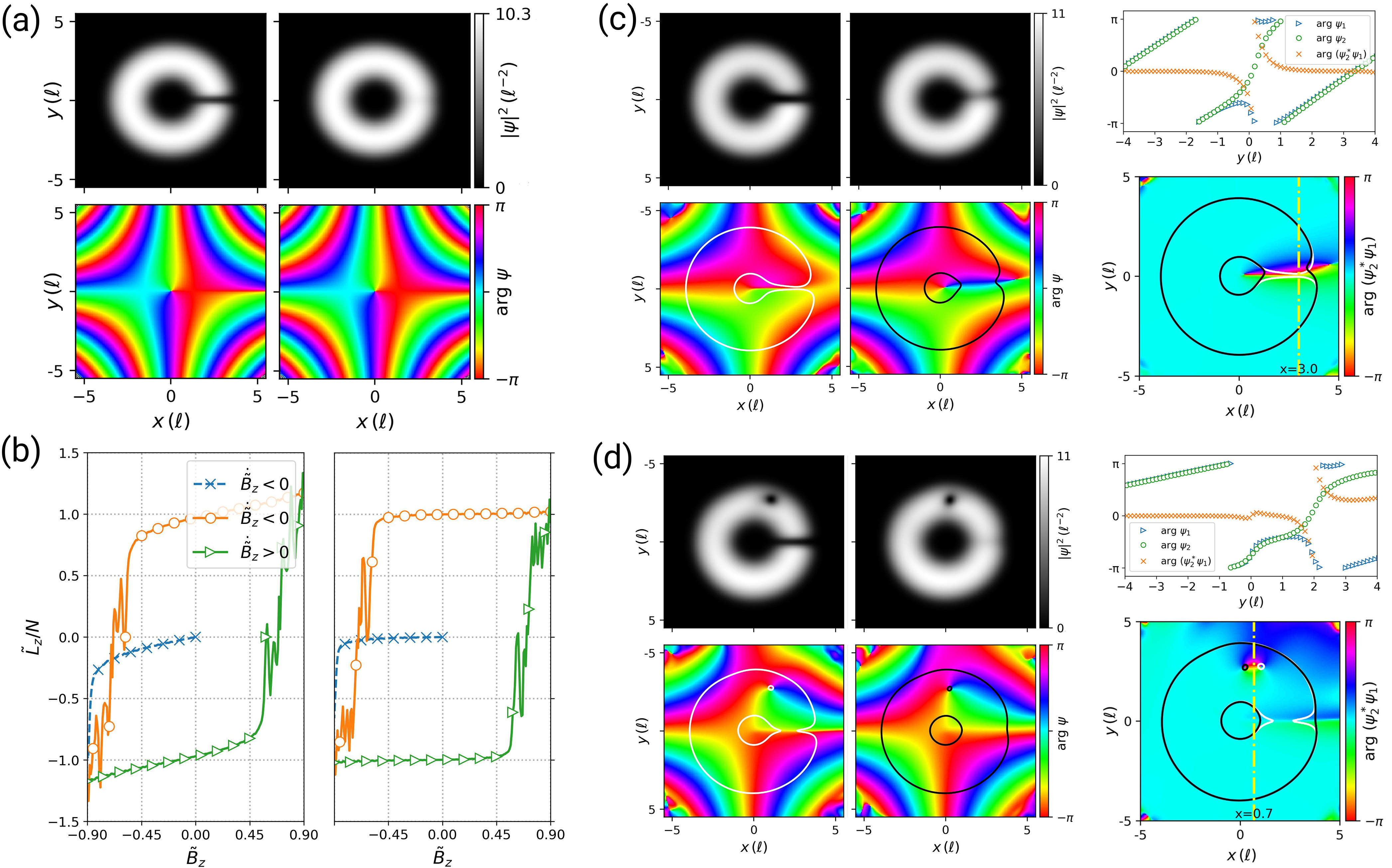}
 	\caption{ Spinor BEC with linear coupling $\tilde\nu=1$ subjected to the adiabatic variation of the synthetic,  perpendicular magnetic field $\tilde B_z=\pm\tilde B_z^0(1-2 t/T)$ with a weak-link potential in one spin component. (a) Ground state at  $\tilde B^0_z= 0.9$. Left and right panels correspond to spin components $\psi_\uparrow$ and $\psi_\downarrow$, respectively. (b) Hysteresis loops. (c) Density, phase, and relative phase of an intermediate state at $\tilde B_z=-0.58$. The relative phase shows a domain wall connecting vortices in different spin components; the 1D phase profiles along the dot-dashed line are shown in the top-right panel.  (d) Same as (c) for lower magnetic field $\tilde B_z=0.6$, where the vortices have entered the density profiles. Solid lines indicate density isocontours and circles indicate vortices. } 
 	\label{fig:nu1}
 \end{figure*}
 Figure \ref{fig:WLsy} illustrates the influence of varying parameters in the adiabatic processes followed by the system considered in Fig. \ref{fig:sWL}, keeping the total number of particles, and so the healing length,  $\tilde \xi\approx 0.3$, constant. The top row shows the hysteresis loops for different azimuthal widths  $s_y$ of the weak link and fixed field variation rate; while short azimuthal barriers $s_y/\xi\ll 1$ gives rise to large hysteresis loops, long azimuthal barriers $s_y/\xi\ge 1$ break down metastability and produce vanishing loops. On the other hand, the bottom row represents the hysteresis loop dependence on the velocity $\dot B_z$ of the adiabatic ramp for fixed azimuthal width; slow velocities (left-most panel) allow the system to transit through the intermediate winding number $m=0$ (thus making transitions $m\to m\pm 1$), which is otherwise passed by (with transitions $m\to m\pm 2$) at higher velocities.  
 As net effect, the latter velocities increase the area of the hysteresis loop in a so-called dynamical hysteresis process, by means of which the system relaxation into a different metastable state is delayed \cite{Broner1997}.

 \begin{figure*}[t]
 	\centering
 	\includegraphics[width=1.0\linewidth]{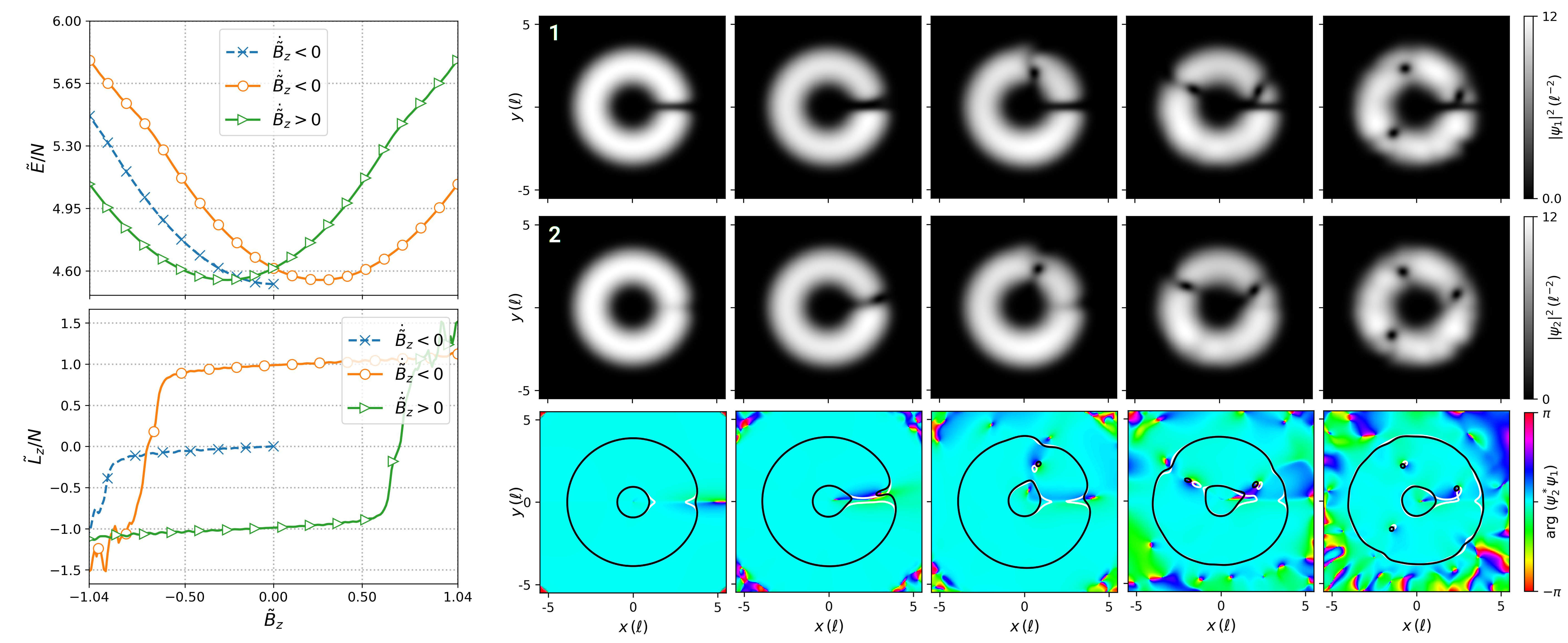}
 	\caption{ Same process as in Fig. \ref{fig:nu1} but with a higher linear coupling  $\tilde\nu=2$. Whole-condensate energy (top) and  angular momentum (bottom) per particle are shown on the left. On the right, density (with labels 1 and 2 indicating spin components $\psi_\uparrow$ and $\psi_\downarrow$) and relative-phase profiles correspond to magnetic field values, from left to right, $\tilde B_z= 1.04,\,-0.66,\,-0.7,\,-0.74,\,-0.8$. } 
 	\label{fig:nu2}
 \end{figure*}
 \subsection{Hysteresis in spinor condensates}

Along with the relative phase $\varphi$, new energy $\nu$ and length $\ell_\nu$ scales are introduced in spinor systems with respect to scalar condensates. These quantities play a major role in driving the  dynamics, mainly when the system is subjected to potentials that explicitly break the symmetry between spin components, as it is the case that we will focus on. On the other hand, spin-symmetric potentials do not produce a qualitative different scenario from scalar condensates, since phase-locked spin components can be expected along adiabatic paths, with the caveat that spin-density (instead of total density) modulations can give rise to instabilities and eventually drive phase slips  \cite{Beattie2013}.

As stated in Eq. (\ref{eq:Hspinor}), our setup includes a weak link operating only in one of the spin components, but it has otherwise equal parameters as in the scalar condensates presented before.
As we show next, this asymmetry between spin components translates, in general, into a different vortex dynamics that originates domain walls of the relative phase \cite{Son2002, Farolfi2020,Kas2004}. These structures connect vortices of equal charge in different spin components, and present characteristic relative-phase jumps $\Delta \varphi=2\pi$ across the length $\ell_\nu$. They can drag both spin components into respective hysteresis loops.

Besides the perpendicular magnetic fields that relate vortices and flux, we will consider in-plane magnetic fields to show a different kind of hysteresis loops related to phase transitions in spin-orbit-coupled systems.

\subsubsection{Perpendicular magnetic field}
To illustrate the typical dynamics, we consider an initial spinor ground state at perpendicular magnetic field $\tilde B_z^0=0.9$ and linear coupling $\tilde \nu=1$, as shown in Fig. \ref{fig:nu1}(a), from which we get the characteristic length $\tilde \ell_\nu= 0.5$; since the rest of parameters do not change with respect to the scalar condensates, we get the ratio $\ell_\nu/\xi=\sqrt{g n/(4\nu)}\sim 1.7$ between healing length and linear coupling length, indicating the lower energy associated with excitations in the relative phase.
Figure \ref{fig:nu1}(b) shows the resulting hysteresis loops followed by both spin components in an adiabatic process of total time $T=180\,\hbar/\nu$. As can be seen at intermediate times, in panel (c)  for $\tilde B_z=-0.58$ and in panel (d) for $\tilde B_z=-0.6$, the double phase-slip process is driven by the appearance of domain walls of the relative phase, whose 1D profiles are well described by 1D Josephson-vortex structures \cite{Kaurov2005}
\begin{align}
\psi_{1,2}(y)&=e^{ik_0 y}\sqrt{n_0}\,\times\nonumber\\ \quad&\left[\tanh\left(\frac{y-y_0}{\ell_\nu}\right)\pm i\, \sqrt{1-\frac{4\nu}{gn_0}}\,\mbox{sech}\left(\frac{y-y_0}{\ell_\nu}\right)\right],
\label{eq:JV}
\end{align}
where $n_0$ and $k_0$ are the asymptotic density and wave number, respectively, away from the vortex core at $y=y_0$.
In them, the relative phase variation is accompanied by a density depletion in the rings (discernible in the figure for the component without weak link) that facilitates the transit of vortices.
These structures exist for linear coupling values $4\nu<gn_0$, or, in other words, for length ratios
$\ell_\nu/\xi>1$. Higher linear couplings suppress Josephson vortices, and other topological defects, such as dark solitons, have to provide the path for vortices in the phase slips. The scenario is confirmed by our numerical simulations; for instance, for the same parameters of Fig. \ref{fig:nu1} but linear coupling $\tilde\nu=4$ (not shown), where the length ratio is $\ell_\nu/\xi\sim 0.8$, we do not find domain walls in the relative-phase profiles and the dynamics evolve in both components through overlapping vortex cores. In this case, the spin components are essentially phase-locked, $\varphi({\bf r})\approx 0$.

Therefore, two asymptotic regimes can be distinguished: when $\nu\ll g n$ the length of the domain walls is large and their energy small(which scales with $\sqrt{\nu}$) \cite{Son2002}, so that it is difficult to drag the second component into a hysteresis loop by making vortex pairs; on the other hand for $\nu\gg g n$ domain walls are not available and excitations of the total density, as it is the case in scalar condensates, are necessary to drive the phase slips in the component without weak link. In the former regime, for $\nu \rightarrow 0$ the dynamics would approach the case of two independent scalar condensates, one with weak link that generates hysteresis loops, and one without weak link that does not present changes in the magnetic flux. Conversely, the high linear-coupling limit would approach the dynamics of a single scalar condensate with weak link. The intermediate regime with $\nu\leq g n$, as illustrated in Fig. \ref{fig:nu1}, and also in Fig. \ref{fig:nu2} with $\tilde \nu=2$ and much shorter domain walls, provides a better way of achieving smooth hysteresis loops. Whereas larger domain walls introduce fast oscillations of greater amplitude during the plase slip event associated with the motion of domain walls, which, non-simultaneously for both spin components, drag (solitonic) vortices into the bulk ring.

Figure 6 depicts the paths followed by the total energy per particle (top left panel) and the total angular momentum per particle (bottom left) in response to the slow variation of the synthetic magnetic field. As can be seen in the right panels, that show a selected sequence of density and relative-phase profiles belonging to the path initiated at $\tilde B_z^0=1.04$, short domain walls joining half vortices (with cores identified by small black and white circles in the relative-phase plots) produce the almost simultaneous entry or exit of vortices, and with them corresponding phase slips, in the ring.

\subsubsection{Spin-orbit coupling: in-plane magnetic field}
\label{sect:soc}
\begin{figure}[t]
	\centering
	\includegraphics[width=\linewidth]{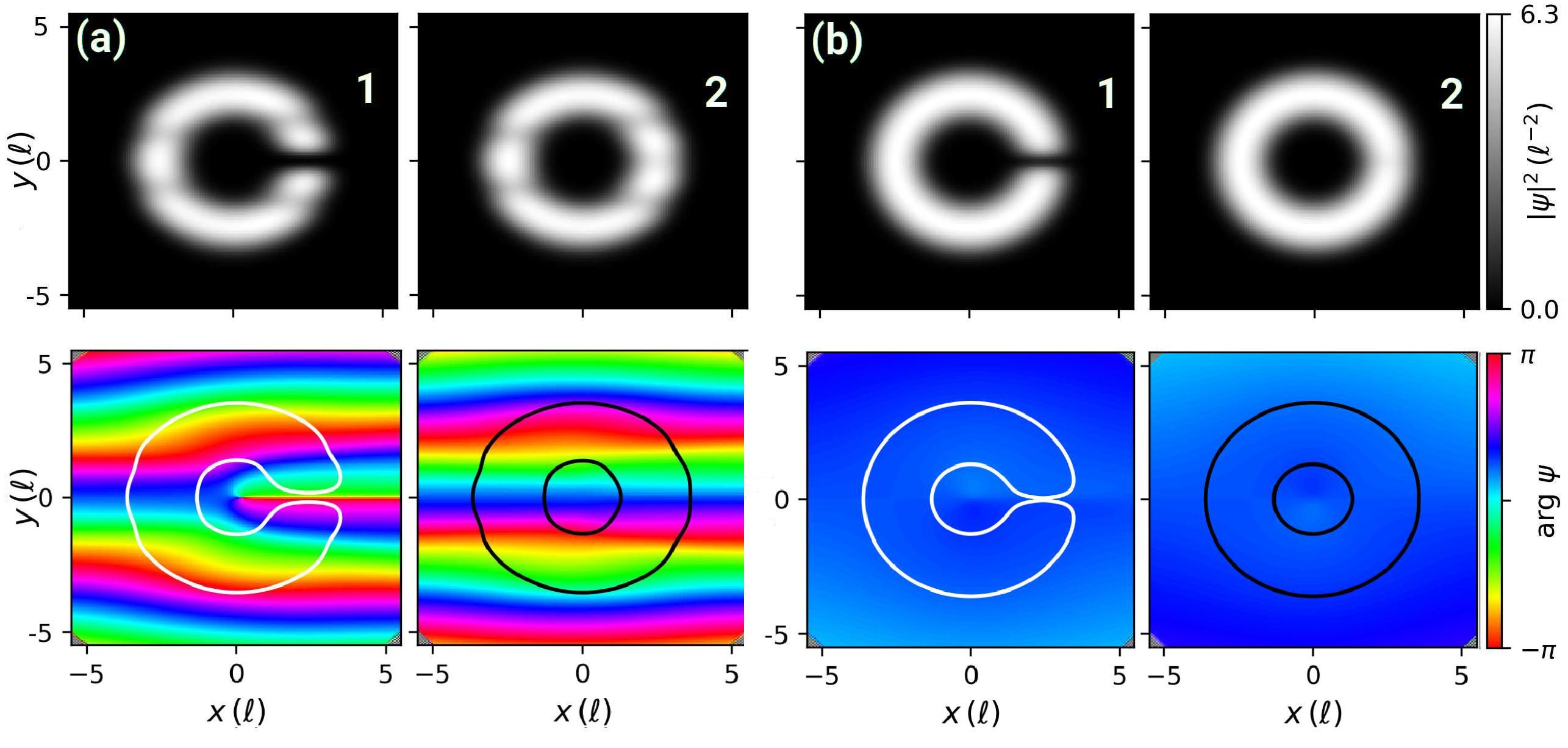}
	\caption{ Ground states of a spinor condensate in the presence of spin orbit coupling with $\tilde B_x=-3.6$ (a) and $\tilde B_x=-0.4$ (b), for fixed linear coupling $\tilde\nu=1$ and total particle number fulfilling $gN/R^2=0.48\nu$.  } 
	\label{fig:SOCnu1}
\end{figure}

In-plane magnetic fields present more difficulties for the generation of hysteresis loops, since phase transitions (see, for instance, Ref. \cite{Li2015}) can be found along the paths of metastable states for varying magnetic field.  In such paths, the spectrum of the system changes, so that different states with no definite winding number are energetically or dynamically favorable to make transitions to. In this situation, hysteresis is not associated with flux variation through loops in the ring plane (since this flux is always zero); instead, it is related to states that can keep metastability around phase transition borders. A similar hysteresis phenomenon has been recently realized in coherently coupled condensates around a (simulated) ferromagnetic phase transition \cite{Cominotti2023}.
Figure \ref{fig:SOCnu1} presents an example of ground states for different phases, stripe and single minimum, obtained in a system with $\tilde \nu=1$ and magnetic field values $\tilde B_x=-3.6$ and $-0.4$, respectively; the particle number has been halved with respect to the previous cases to enhance the contrast in the density stripes.  The former state shows a half vortex (a vortex in just one component) at the center of the  ring with slit, whereas density modulations appear in both components. On the other hand, the state at low magnetic field has an almost flat phase profile and (apart from the slit) homogeneous density profile. An adiabatic path joining these states, that belong to different phases of the system and are far apart in the parameter space, cannot preserve the metastability of the initial state, and gives rise to complex transient configurations.

\begin{figure}[t]
	\centering
	\includegraphics[width=\linewidth]{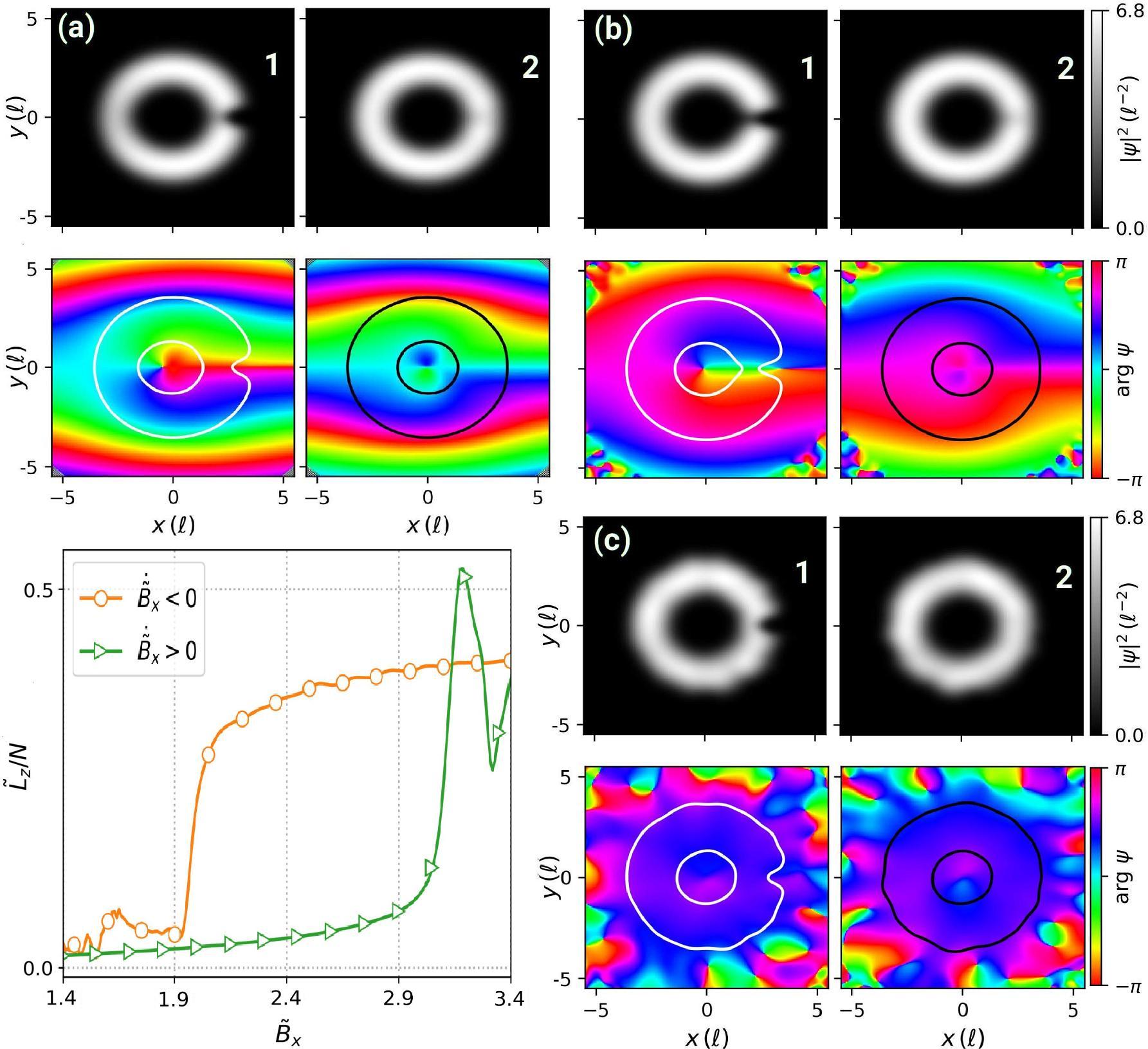}
	\caption{ Total angular momentum per particle (bottom left) in the presence of spin orbit coupling with  the same parameters of Fig. \ref{fig:SOCnu1}. The density and phase profiles (a), (b), (c), are snapshots of the orange path (with open circles) taken at magnetic field values $\tilde B_x=3.4,\,2.2,\,1.4$, respectively. } 
	\label{fig:SOCloop}
\end{figure}

Alternatively, adiabatic paths that preserve the metastable states, hence producing hysteresis loops, are possible when they transit in the neighborhood of a phase transition, that is, paths that do not reach deep inside the neighbor phases. Figure \ref{fig:SOCloop} illustrates this scenario in a system with a less intrusive weak link that affects only the outer part of the ring, with $x_0=1.28\, R$ in Eq. (\ref{eq:W}). Following the orange curve (with open circles) for the total angular momentum per particle, the initial ground state at $\tilde B_x^0=3.4$  presents a half vortex in one spin component, panel (a), which becomes part of a vortex dipole at intermediate field values,  panel (b), and eventually annihilates, panel (c), to produce zero angular momentum. As a result, the observed hysteresis loop in the total angular momentum chart corresponds only to the entry or exit of this half vortex. The other spin component (labeled as number 2) does not change qualitatively its angular momentum, which differs from zero only in oscillations of small amplitude, and so it is not dragged into a hysteresis loop.

\section{Conclusions}
\label{sec:conclusions}

We have explored the generation of hysteresis loops in pseudo-spin-1/2 Bose-Einstein condensates  in the presence of varying synthetic gauge fields that generalize previous studies on rotating scalar systems. Weak-link potentials, which operate in just one of the system spin components, are demonstrated to drive the current-carrying metastable states into phase-slip events that produce hysteresis loops in both spin components. 
Our results show that domain walls of the relative phase are dynamically generated between half vortices, by means of which passages among different current states are possible. The system dynamics depend on the direction of the synthetic magnetic field: whereas perpendicular magnetic fields give rise, typically, to large hysteresis loops that involve magnetic flux changes and are built on metastable states differing in two units of the winding number $|\Delta m|=2$, in-plane (spin-orbit-coupling) fields are based on, and restricted by, the presence of phase transitions and exhibit small hysteresis loops between contiguous $|\Delta m|=1$ winding numbers.

Due to the dynamical stability of domain walls in the relative phase \cite{Son2002,Farolfi2020}, these results, which are within reach of current experimental research, suggest a way of winding and unwinding the phase of the condensates in a non-dissipative way, differently to the case of scalar systems.
Further work on the properties of stationary domain walls associated with the presence of weak links are beyond the present work and will be reported elsewhere.

\begin{acknowledgments}
S.J., J.X., Q.J., and H.Q. acknowledge support by the National
Natural Science Foundation of China (Grant No. 11402199),
the Natural Science Foundation of Shaanxi Province (Grants
No. 2022JM-004 and No. 2018JM1050), and the Educa-
tion Department Foundation of Shaanxi Province (Grant No.
14JK1676).
\end{acknowledgments}

\bibliographystyle{apsrev4-1}
\bibliography{gauge_vortex}

\end{document}